\documentclass[aps,prstab,twocolumn,superscriptaddress,
showpacs,amsmath,amssymb,floatfix,letterpaper,bibnotes]{revtex4} 

\usepackage{graphicx}

\newcommand{\lae}{\lower 2pt \hbox{$\, \buildrel {\scriptstyle <}\over {\scriptstyle\sim}\,$}}

\begin{document} 

\title{Symmetric achromatic low-beta collider interaction region design concept} 

\author{V.S.~Morozov}
\author{Ya.S.~Derbenev} 
\author{F.~Lin}
\affiliation{Thomas Jefferson National Accelerator Facility, Newport News, Virginia 23606, USA}

\author{R.P.~Johnson}
\affiliation{Muons, Inc., Batavia, Illinois 60510, USA}

\date{August 16, 2012} 

\begin{abstract} 

We present a new symmetry-based concept for an achromatic low-beta collider interaction region design. 
A specially-designed symmetric Chromaticity Compensation Block (CCB) induces an angle spread in the 
passing beam such that it cancels the chromatic kick of the final focusing quadrupoles. 
Two such CCBs placed symmetrically around an interaction point allow simultaneous compensation of the 
1st-order chromaticities and chromatic beam smear at the IP without inducing significant 2nd-order 
aberrations to the particle trajectory. We first develop an analytic description of this approach 
and explicitly formulate 2nd-order aberration compensation conditions at the interaction point. 
The concept is next applied to develop an interaction region design for the ion collider ring of 
an electron-ion collider. We numerically evaluate performance of the design in terms of momentum 
acceptance and dynamic aperture. The advantages of the new concept are illustrated by comparing it 
to the conventional distributed-sextupole chromaticity compensation scheme.

\end{abstract} 

\pacs{29.20.db, 29.27.Bd, 41.75.-i, 41.85.Gy} 
\keywords{low-beta collider, interaction region design, chromatic beam smear, chromaticity compensation, 
momentum acceptance, dynamic aperture}

\maketitle 

\section{Introduction}
In order to achieve the highest possible luminosity in a collider~\cite{des_rep12,erhic_zdr}, 
the colliding beams must be focused to a small spot at the Interaction Point (IP). This tight focusing 
is unavoidably accompanied by a large transverse beam extension before the beam enters the Final 
Focusing Block (FFB). The size of the required beam extension is determined by the desired degree of 
beam squeezing at the IP and the focal length of the FFB, which is closely related to the space between 
the IP and the nearest focusing quadrupole. The larger this distance, the greater the beam extension 
must be. Since the FFB's focal length depends on the particle's momentum, the large beam size inside 
the FFB leads to a large correlation between the particle's phase advance and its 
momentum~\cite{handbook_ape}. 

The problem with such a correlation is two-fold. First of all, it induces a large chromatic betatron 
tune spread. Since the available betatron tune space is limited by the beam resonances, the chromatic 
betatron tune dependence limits the ring's momentum aperture. Secondly, the chromatic dependence of 
the focal length causes chromatic beam smear at the IP, which can even dominate over the 
beam size due to the emittance, significantly increasing the beam spot size at the IP and resulting 
in luminosity loss. Thus, an Interaction Region (IR) design must employ sextupole magnets to compensate 
the chromatic effects~\cite{nosochkov92,kekb_des_rep,raimondi01,alexahin11}. However, the nonlinear 
sextupole fields generate 2nd- and higher-order aberrations to the particle trajectory at the IP, 
introduce nonlinear phase advance and limit the ring's dynamic aperture. Compensation of these 
nonlinear effects is one of the main challenges of an IR and, even more generally, 
collider ring design. A commonly used chromaticity compensation technique is to install same-strength 
sextupoles in pairs with $-I$ (minus the identity) transformation between them to cancel their nonlinear 
kicks~\cite{handbook_ape,lee99}. However, this approach does not treat all of the sextupole-induced 
2nd-order effects. Besides, organizing $-I$ transformations for the sufficient number of sextupole pairs 
for simultaneous compensation of the horizontal and vertical chromaticities is difficult. The concept 
described below not only includes all of the $-I$ transformation benefits but goes further by compensating 
all of the 2nd-order effects together with the 2nd-order beam smear at the IP. It also offers a more 
compact compensation scheme.

The IR design approach presented in this paper involves installation of a dedicated Chromaticity 
Compensation Block (CCB) next to the FFB~\cite{derbenev07,wang09,morozov_ipac11,lin_ipac12}. 
In the CCB, certain symmetries 
of the beam orbital motion and dispersion are created using a symmetric arrangement of dipoles and 
quadrupoles. Two such CCBs are placed symmetrically around the IP. The symmetries of the beam orbital 
motion and dispersion combined with a symmetric quadrupole and sextupole arrangement in the CCBs 
then allow simultaneous compensation of the 1st-order chromaticities and chromatic beam smear 
at the IP without inducing significant 2nd-order aberrations and therefore helping 
preserve the ring's dynamic aperture.

\begin{figure}[ht!]
\includegraphics[width=\columnwidth]{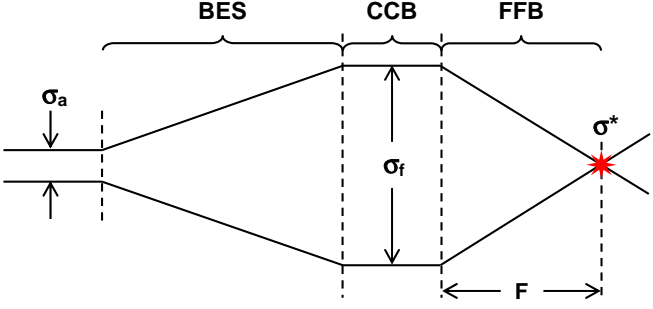}
\caption{Schematic IR layout consisting of Beam Extension Section (BES), Chromaticity Compensation 
Block (CCB), and Final Focusing Block (FFB).} 
\label{fig:IRscheme1}
\end{figure}

A schematic IR layout is shown in Fig.~\ref{fig:IRscheme1}. The Beam Extension Section (BES) first 
expands the beam from its regular size in the arcs to the size required for final focusing. 
The beam next passes through the CCB, which creates in it an angle spread negatively correlated 
with the chromatic kick of the FFB, so that the FFB's chromatic effect is cancelled. The FFB then 
focuses the beam to the required spot size at the IP. The IR design is mirror-symmetric with respect 
to the IP.

\section{Analytic description}
In general, compensation of chromatic effects of a focusing lattice requires introduction 
of dipole and sextupole magnets along the lattice. The dipole field changes the beam reference 
orbit and creates dispersion, i.e. orbit dependence on the particle energy. The sextupole 
field then introduces an additional energy-dependent focusing strength to counter the 
original chromatic effect of the quadrupoles.

The particle motion in the horizontal $x$ and vertical $y$ planes about the reference orbit 
is described by the following equations~\cite{courant58}:
\begin{equation}
\begin{split}
& x''+(K^2-n)x=K\delta+f_x, \\
& y''+ny=f_y,
\end{split}
\label{eq:eqs_of_motion}
\end{equation}
where the prime $'$ denotes a derivative with respect to the longitudinal coordinate $s$, 
$K \equiv 1/\rho$ is the curvature of the bent reference orbit, 
$n \equiv (\partial B_y/\partial x)/(p/|e|)$ is the normalized quadrupole strength, and 
$\delta \equiv \Delta p/p$ is the particle's relative momentum deviation from the reference 
value. The force terms $f_x$ and $f_y$ in Eq.~(\ref{eq:eqs_of_motion}) up to the 2nd order in 
$\delta$, $x$, and $y$ are given by
\begin{equation}
\begin{split}
& f_x = -n \delta x + n_s (x^2-y^2), \\
& f_y = n \delta y - 2 n_s x y,
\end{split}
\label{eq:force_terms}
\end{equation}
where $n_s \equiv (\partial^2 B_y/\partial x^2)/(p/|e|)$ is the normalized sextupole strength. 
The quantities $K$, $n$, $n_s$, and, therefore, $f_x$ and $f_y$ are functions of $s$. 

The particle trajectory in the horizontal plane can be represented as a superposition of 
the dispersive component $D\delta$ and the betatron component $x_b$: 
\begin{equation}
x = D \delta + x_b,
\label{eq:traj_comp}
\end{equation}
where the dispersion function $D \equiv \partial x/\partial \delta$ satisfies the equation 
\begin{equation}
D''+(K^2-n)D=K.
\label{eq:disp}
\end{equation}
Then the evolution of the betatron component is determined by the following equations: 
\begin{equation}
\begin{split}
x_b'' & +(K^2-n)x_b=f_x =-(n-2D n_s)\delta x_b \\ 
& +D(D n_s-n)\delta^2+n_s(x_b^2-y^2), \\ 
y'' & +ny=f_y=(n-2D n_s)\delta y-2n_s x_b y. 
\end{split} 
\label{eq:bet_comp} 
\end{equation} 

Sextupoles are usually placed in the ring arcs to compensate the energy dependence of 
the global betatron tunes. In that case, the chromaticity compensating sextupoles need 
to be configured to satisfy only two conditions associated with the $\delta x_b$ and 
$\delta y$ terms on the right hand side of Eq.~(\ref{eq:bet_comp}), i.e. the integral 
effect of these terms over one revolution must be zero. The influence of the other three 
terms in Eq.~(\ref{eq:bet_comp}) on the particle amplitude and global phase advance is 
zero when averaged over many revolutions. However, local chromatic effects in short 
sections or even at specific points like the IP where the beam experiences 
a strong transverse compression are generally not compensated on average in one turn. 
Compensation of these effects is a much more complicated problem, which requires taking 
into account all terms in Eq.~(\ref{eq:bet_comp}). As one can see from these equations, 
the sextupole field component and dispersion give rise to the undesirable quadratic 
aberrations:   
\begin{equation}
\begin{split}
& \Delta x'' \propto n_s [(D\delta)^2+x_b^2-y^2]-D n \delta^2, \\ 
& \Delta y'' \propto -2n_s x_b y. 
\end{split}
\label{eq:quad_aberr}
\end{equation} 
Compensating the impact of these terms on the beam size at the IP, 
where the beam size is supposed to be extremely small, requires satisfying five 
different conditions in total. In order to explicitly state these conditions, 
we use the analytic perturbation approach below.

We represent the betatron motion as 
\begin{equation}
\begin{split}
x_b & = x_1+\tilde{x}, \\ 
y & = y_1+\tilde{y}. 
\end{split}
\label{eq:pert}
\end{equation} 
where the unperturbed linear parts $x_1$ and $y_1$ satisfy the homogeneous 
equations:
\begin{equation}
\begin{split}
& x_1'' + (K^2-n)x_1 = 0, \\ 
& y_1'' + n y_1 = 0,  
\end{split}
\label{eq:1st_order}
\end{equation} 
while the perturbed trajectory components are given by
\begin{equation}
\begin{split}
& \tilde{x}'' + (K^2-n)\tilde{x} = f_x, \\ 
& \tilde{y}'' + n \tilde{y} = f_y.  
\end{split}
\label{eq:pert_comp}
\end{equation} 
We next express $x_1$ and $y_1$ as linear combinations of the two linearly-independent 
solutions $u(s)$ and $v(s)$ of Eq.~(\ref{eq:1st_order}):
\begin{equation}
\begin{split}
x_1 & = a_x u_x(s)+b_x v_x(s), \\ 
y_1 & = a_y u_y(s)+b_y v_y(s).  
\end{split}
\label{eq:lin_comb}
\end{equation} 
We then use the parameter variation technique to solve for the perturbed part
\begin{equation}
\begin{split}
\tilde{x} & = \tilde{a}_x(s) u_x(s)+\tilde{b}_x(s) v_x(s), \\ 
\tilde{y} & = \tilde{a}_y(s) u_y(s)+\tilde{b}_y(s) v_y(s).  
\end{split}
\label{eq:par_var}
\end{equation} 
Using the Cramer's rule, the formal solution for the perturbed part can be written as
\begin{equation}
\begin{split}
\tilde{x}(s) & = \frac{1}{W_x} [v_x \int_0^s u_x f_x ds - u_x \int_0^s v_x f_x ds], \\ 
\tilde{y}(s) & = \frac{1}{W_y} [v_y \int_0^s u_y f_y ds - u_y \int_0^s v_y f_y ds],  
\end{split}
\label{eq:pert_sol}
\end{equation} 
where $W_x \equiv u_x v_x'-u_x'v_x$ and $W_y$ are the Wronskians of the $x$ and $y$ 
basis functions $u$ and $v$.

Suppose that $u$ and $v$ are chosen such that at the IP $u^\ast=0$ and 
$v'^\ast=0$, then the perturbed trajectory components at the IP are given 
by
\begin{equation}
\begin{split}
\tilde{x}^\ast & = \frac{1}{W_x} v_x^\ast \int_0^\ast u_x f_x ds 
= -\frac{1}{u_x'^\ast} \int_0^\ast u_x f_x ds, \\ 
\tilde{y}^\ast & = \frac{1}{W_y} v_y^\ast \int_0^\ast u_y f_y ds 
= -\frac{1}{u_y'^\ast} \int_0^\ast u_y f_y ds. 
\end{split}
\label{eq:pert_at_ip}
\end{equation} 
By performing the integration in Eq.~(\ref{eq:pert_at_ip}) over the whole beam 
path from the start of the beam extension after the arc to the IP, 
we obtain the perturbation of the particle transverse position at the IP caused 
by the aberrations from all of the involved bends, quadrupoles, and sextupoles.

To find the next order correction to the unperturbed solution $x_1$ and $y_1$, 
in accordance with the iteration method, we plug the 1st-order unperturbed solution 
$x=D\delta+x_1$ and $y=y_1$ into the right-hand side of Eq.~(\ref{eq:pert_comp}), 
obtaining the following equations for the 2nd-order correction:
\begin{equation}
\begin{split}
\tilde{x}_2'' & -n \tilde{x}_2= -(n-2D n_s)\delta x_1 \\
& +D(D n_s-n)\delta^2 +n_s(x_1^2-y_1^2), \\ 
\tilde{y}_2'' & +n \tilde{y}_2= (n-2D n_s)\delta y_1-2n_s x_1 y_1, 
\end{split} 
\label{eq:pert_2nd_order} 
\end{equation} 
where we neglected the term $K^2 x$ on the left-hand side of the $x$ equation, 
since the effect of the orbit curvature on beam focusing at high energy is very small 
compared to that of the quadrupole field. Using Eq.~(\ref{eq:pert_at_ip}), the 2nd-order 
correction to the particle's unperturbed trajectory at the IP is given by
\begin{equation}
\begin{split}
\tilde{x}_2^\ast = & -\frac{1}{u_x'^\ast} \int_0^\ast u_x [-(n-2D n_s)\delta x_1 \\ 
& +D(D n_s-n)\delta^2 +n_s(x_1^2-y_1^2)] ds, \\ 
\tilde{y}_2^\ast = & -\frac{1}{u_y'^\ast} \int_0^\ast u_y [
(n-2D n_s)\delta y_1-2n_s x_1 y_1] ds. 
\end{split}
\label{eq:2nd_order_at_ip}
\end{equation} 

Consider an almost parallel beam after the beam extension. The function $u$ then describes 
the dominant (``cos''-like) parallel component of the trajectory while the orthogonal 
solution $v$ is associated with the small remaining angular spread (``sin''-like trajectory). 
Then, neglecting the angular divergence, one can use Eq.~(\ref{eq:lin_comb}) to approximate 
$x_1$ and $y_1$ in Eq.~(\ref{eq:2nd_order_at_ip}) as
\begin{equation}
\begin{split}
x_1 & \approx a_x u_x(s), \\ 
y_1 & \approx a_y u_y(s),  
\end{split}
\label{eq:1st_order_approx}
\end{equation} 
obtaining for the 2nd-order trajectory correction at the IP
\begin{equation}
\begin{split}
\tilde{x}_2^\ast \approx & -\frac{1}{u_x'^\ast} \int_0^\ast u_x [-(n-2D n_s)\delta a_x u_x \\ 
& +D(D n_s-n)\delta^2 +n_s(a_x^2 u_x^2-a_y^2 u_y^2)] ds, \\ 
\tilde{y}_2^\ast \approx & -\frac{1}{u_y'^\ast} \int_0^\ast u_y [(n-2D n_s)\delta a_y u_y \\ 
& -2n_s a_x u_x a_y u_y] ds. 
\end{split}
\label{eq:2nd_order_at_ip_approx}
\end{equation} 
Clearly, aberration compensation requires
\begin{equation}
\begin{split}
\tilde{x}_2^\ast = 0, \\ 
\tilde{y}_2^\ast = 0. 
\end{split}
\label{eq:2nd_order_comp_req}
\end{equation} 
Equation~(\ref{eq:2nd_order_at_ip_approx}) then leads to the following five compensation 
conditions:
\begin{equation}
\begin{split}
& (1)\hspace{0.5cm} 2\int_0^\ast D n_s u_x^2 ds = \int_0^\ast n u_x^2 ds, \\ 
& (2)\hspace{0.5cm} 2\int_0^\ast D n_s u_y^2 ds = \int_0^\ast n u_y^2 ds, \\ 
& (3)\hspace{0.5cm} \int_0^\ast D (D n_s - n) u_x ds = 0, \\ 
& (4)\hspace{0.5cm} \int_0^\ast n_s u_x^3 ds = 0, \\ 
& (5)\hspace{0.5cm} \int_0^\ast n_s u_x u_y^2 ds = 0. 
\end{split}
\label{eq:2nd_order_comp_cond_1}
\end{equation} 
Conditions (1) and (2) in Eq.~(\ref{eq:2nd_order_comp_cond_1}) are the usual conditions for 
compensation of the chromatic spread. Conditions (3)-(5) are the requirements for 
compensation of the 2nd-order dispersion effect and sextupole beam smear.

In the above analysis of the perturbed trajectory component, we so far ignored contribution 
of the angular spread, which is associated with the solution $v$. Even though this contribution 
is small compared to that of $u$, it may become dominant after compensation of the main 
aberration terms. Perhaps more importantly, the nonlinear phase advance associated with 
the angular spread may adversely affect the dynamic aperture. Using Eqs.~(\ref{eq:lin_comb}) 
and (\ref{eq:2nd_order_at_ip}) and neglecting the terms of the 2nd order in $v$, contribution 
of the angular spread to the 2nd-order perturbation at the IP can be written as
\begin{equation}
\begin{split}
\Delta\tilde{x}_2^\ast \approx & -\frac{1}{u_x'^\ast} \int_0^\ast u_x [-(n-2D n_s)\delta b_x v_x \\ 
& +2n_s(a_x u_x b_x v_x - a_y u_y b_y v_y)] ds, \\ 
\Delta\tilde{y}_2^\ast \approx & -\frac{1}{u_y'^\ast} \int_0^\ast u_y [(n-2D n_s)\delta b_y v_y \\ 
& -2n_s (a_x u_x b_y v_y + a_y u_y b_x v_x) ] ds. 
\end{split}
\label{eq:2nd_order_at_ip_angle}
\end{equation} 
Compensation of the aberration terms in Eq.~(\ref{eq:2nd_order_at_ip_angle}) imposes five more 
constraints in addition to those in Eq.~(\ref{eq:2nd_order_comp_cond_1}):
\begin{equation}
\begin{split}
& (1)\hspace{0.5cm} 2\int_0^\ast D n_s u_x v_x ds = \int_0^\ast n u_x v_x ds, \\ 
& (2)\hspace{0.5cm} 2\int_0^\ast D n_s u_y v_y ds = \int_0^\ast n u_y v_y ds, \\ 
& (3)\hspace{0.5cm} \int_0^\ast n_s u_x^2 v_x ds = 0, \\ 
& (4)\hspace{0.5cm} \int_0^\ast n_s u_x u_y v_y ds = 0, \\ 
& (5)\hspace{0.5cm} \int_0^\ast n_s u_y^2 v_x ds = 0. 
\end{split}
\label{eq:2nd_order_comp_cond_2}
\end{equation} 

Corrections beyond the 2nd order can be obtained by considering further iterations:
\begin{equation}
\begin{split}
\tilde{x} & = \tilde{x}_2+\tilde{x}_3+\ldots, \\ 
\tilde{y} & = \tilde{y}_2+\tilde{y}_3+\ldots 
\end{split}
\label{eq:higher_order_pert}
\end{equation} 
Thus, the 3rd-order correction terms can be found by integrating the following equations: 
\begin{equation}
\begin{split}
\tilde{x}_3'' & -n\tilde{x}_3 \approx n \delta^2 x_1 + (2D n_s - n)\delta \tilde{x}_2 \\ 
& +2n_s(x_1\tilde{x}_2-y_1\tilde{y}_2)-\frac{1}{6}n''(D\delta+x_1)^3 \\
& +n_o [(D\delta+x_1)^3-3(D\delta+x_1)y_1^2], \\ 
\tilde{y}_3'' & +n\tilde{y}_3 \approx -n \delta^2 y_1 - (2D n_s - n)\delta \tilde{y}_2 \\ 
& -2n_s(x_1\tilde{y}_2+y_1\tilde{x}_2)+\frac{1}{6}n''y_1^3 \\
& +n_o [y_1^3-3(D\delta+x_1)^2 y_1], 
\end{split}
\label{eq:3rd_order}
\end{equation} 
where we introduced the octupole field component $n_o \equiv (\partial^3 B_y/\partial x^3)/(p/|e|)$. 
The higher-order terms should be studied to determine their effects on the beam smear at the 
IP and the ring's dynamic aperture. Equation~(\ref{eq:3rd_order}) can be integrated 
similarly to Eq.~(\ref{eq:pert_2nd_order}); however, a complete analytic analysis clearly becomes 
rather cumbersome even in this 3rd-order case and one must employ a numerical approach. Nevertheless, 
one can still use Eq.~(\ref{eq:3rd_order}) to identify the symmetries helpful in compensating the 
higher-order aberrations to guide the numerical studies.

\section{IR linear optics design}
Equation~(\ref{eq:2nd_order_comp_cond_1}) is an analytic representation of the interaction 
region design requirements up to the 2nd order in effects on the particle trajectory. By imposing 
certain symmetries of the orbital motion, dispersion and magnetic field components with respect 
to the center of the CCB, namely,
\begin{equation}
\begin{split}
& u_x(s)=\mp u_x(-s), \\ 
& u_y^2(s)=u_y^2(-s), \\ 
& D(s)=\pm D(-s), \\ 
& n(s)=n(-s), \\ 
& n_s(s)=\pm n_s(-s), 
\end{split}
\label{eq:symm_cond}
\end{equation} 
the five conditions of Eq.~(\ref{eq:2nd_order_comp_cond_1}) are reduced to just the first two, 
i.e. the linear chromaticity compensation conditions. While creating chromatic kick, the CCB 
does not generate the nonlinear aberrations associated with the 2nd-order effects of the 
dispersion and transverse beam sizes in the sextupole fields. All these terms corresponding 
to conditions (3)-(5) of Eq.~(\ref{eq:2nd_order_comp_cond_1}) are automatically compensated 
inside the CCB due to the special optics design and lattice symmetry. 
Equation~(\ref{eq:2nd_order_comp_cond_1}) ignores the effect of the beam angular spread, 
which is admittedly small since the beam is assumed to be greatly extended and almost parallel 
at the entrance into the CCB. However, this effect may adversely affect the ring's dynamic aperture. 
The 2nd-order terms associated with the beam's angular spread are given by 
Eq.~(\ref{eq:2nd_order_comp_cond_2}). The symmetry conditions in Eq.~(\ref{eq:symm_cond}) do 
not give any advantage in this case. The significance of these and higher-order terms needs 
to be examined in each specific case and any required compensation may be done by introducing 
additional sextupole and higher-order multipole families, perhaps, also with appropriate types 
of symmetry.

We next test our symmetry concept by developing a conceptual IR design~\cite{morozov_ipac11,lin_ipac12} 
for the ion collider ring of the Medium-energy Electron-Ion Collider (MEIC) that is being 
proposed by Jefferson Lab~\cite{des_rep12}. Note that this is a proof-of-principle demonstration 
rather than a final IR design. It is used primarily to validate our symmetry-based chromaticity 
compensation concept. However, due to the modularity of our approach and the fact that 
the concept is independent of a particular FFB design, the IR design presented in this paper can be 
easily optimized to meet any specific nuclear physics detector requirements as discussed 
in~\cite{des_rep12,morozov_ipac12}. 

\begin{figure}[b!]
\includegraphics[width=\columnwidth]{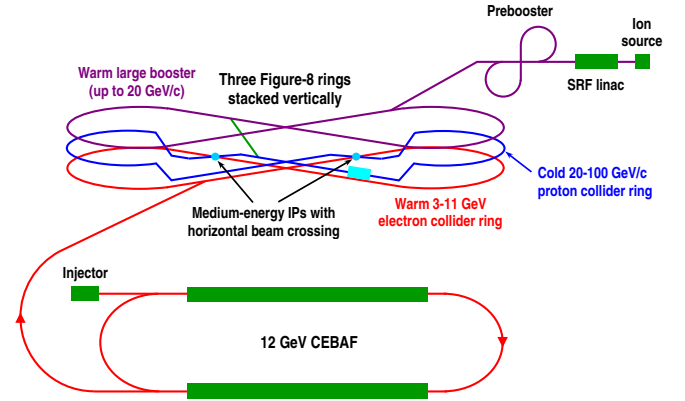}
\caption{Schematic layout of the proposed MEIC complex.} 
\label{fig:MEIClayout}
\end{figure} 

Figure~\ref{fig:MEIClayout} shows a proposed layout of the MEIC complex~\cite{des_rep12}.
The main MEIC parameters are summarized in Table~\ref{tab:MEICpar}. 
As shown in Fig.~\ref{fig:MEIClayout}, the electron and ion collider rings have 
geometrically-matching figure-8 shapes consisting of two $240^\circ$ arcs connected by two straight 
sections crossing each other in the middle at $60^\circ$. There is a total of two IPs: one IP per straight. 
The rings are stacked vertically in the arcs but, at the IPs, the beams cross in the horizontal plane. 
This is done by keeping the electron ring flat and bringing the ion beam to the electron beam's plane 
using vertical chicanes located at the ends of the arcs. 

\begin{table}[t]
\caption{Main MEIC parameters.}
\begin{ruledtabular}
\begin{tabular}{lccc}
Parameter & Unit & Electrons  & Protons \\ \hline
Momentum range & GeV/$c$ & 3-11 & 20-100 \\
Optimization point & GeV/$c$ & 5 & 60 \\
Polarization & \% & $\sim 80$ & $>70$ \\
Collision frequency & MHz &    \multicolumn{2}{c}{748.5}    \\
Particles per bunch & $10^{10}$ & 2.5 & 0.42 \\
Beam current & A & 3.0 & 0.5 \\
$\Delta p/p$ (rms) & $10^{-3}$ & 0.7 & 0.3 \\
Bunch length (rms) & mm & 7.5 & 10 \\
Norm. $x$ emit. (rms) & $\mu$m & 54 & 0.35 \\
Norm. $y$ emit. (rms) & $\mu$m & 11 & 0.07 \\
$\beta_x$ at IP $(\beta_x^*)$ & cm & \multicolumn{2}{c}{10} \\
$\beta_y$ at IP $(\beta_y^*)$ & cm & \multicolumn{2}{c}{2} \\
Collider ring length  & m & \multicolumn{2}{c}{1340} \\
Luminosity per IP & cm$^{-2}$s$^{-1}$ & \multicolumn{2}{c}{$5.6\times 10^{33}$}  \\
\end{tabular}
\end{ruledtabular}
\label{tab:MEICpar}
\end{table}

Since the IR design concepts are similar for the two collider rings, below we focus 
our discussion on the ion collider ring, which is perhaps more challenging due to its 
larger natural linear chromaticities~\cite{des_rep12}. Two IRs are incorporated into 
the ion ring's two straight sections. The IR sections are assumed 
identical and the ring is arranged in a two-fold symmetric way. 
The ring's horizontal $\nu_x$ and vertical $\nu_y$ betatron tunes are set 
to 23.273 and 21.285, respectively, by adjusting the phase advance outside of the IRs.
The ring lattice used in the below simulations is fairly basic; it does not include 
components such as RF cavities, detector solenoids, vertical chicanes, spin control devices, 
injection/extraction elements, etc. However, the details of the ring design outside of the 
IR sections have little impact on the results of this paper and their description can be 
found in~\cite{des_rep12}. Therefore, below we focus our attention on the IR design.

\begin{figure}[t!]
\includegraphics[width=\columnwidth]{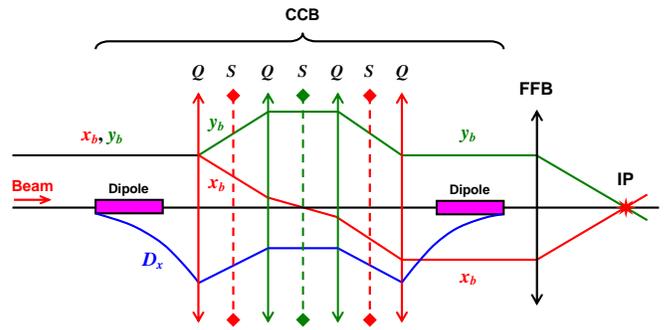}
\caption{A conceptual drawing of an IR with a CCB that has an 
odd-symmetry horizontal betatron trajectory $x_b$, an even-symmetry vertical betatron trajectory $y_b$, 
and an even-symmetry horizontal dispersion $D_x$. It shows a possible arrangement of 
dipoles, quadrupoles $Q$, and sextupoles $S$ that satisfies the magnetic field symmetry 
requirements and produces the desired symmetries of $x_b$, $y_b$, and $D_x$.} 
\label{fig:IRscheme2}
\end{figure}

A conceptual drawing of an IR with a CCB that has an even symmetry of the dispersion and an odd 
symmetry of the horizontal betatron trajectory is shown in Fig.~\ref{fig:IRscheme2}. If one then 
uses an even symmetry of the sextupole fields, conditions (3)-(5) of 
Eq.~(\ref{eq:2nd_order_comp_cond_1}) are automatically satisfied. The two remaining chromaticity 
compensation conditions are satisfied by adjusting the fields of, at least, two sextupole families. 
This can be attained using the difference in the behavior of the horizontal and vertical 
$\beta$-functions.

In Fig.~\ref{fig:IRscheme2}, there are two identical bends at the beginning and at the end 
of the CCB, which generate and then suppress the dispersion while symmetric quadrupole optics 
ensures the appropriate symmetries of the betatron and dispersive orbital components with 
respect to the center of the CCB. When designing the ion interaction region of an electron-ion 
collider, one has to keep in mind that it has to match the footprint of the electron interaction 
region. In the electron ring, since the CCB dipoles are located in regions with large 
$\beta$-function values, their maximum bends must be limited to avoid their emittance-degrading 
impact. In the ion ring, on the other hand, it is advantageous to have strong bends to produce a 
large dispersion thus reducing required sextupole fields and their nonlinear effects. These 
bend requirements are somewhat contradicting. Our solution is to use alternating but still 
symmetric bends in the ion CCB. One can then generate a large dispersion while having the 
flexibility of the footprint geometry.

\begin{figure}[b!]
\includegraphics[height=0.95\columnwidth,angle=-90]{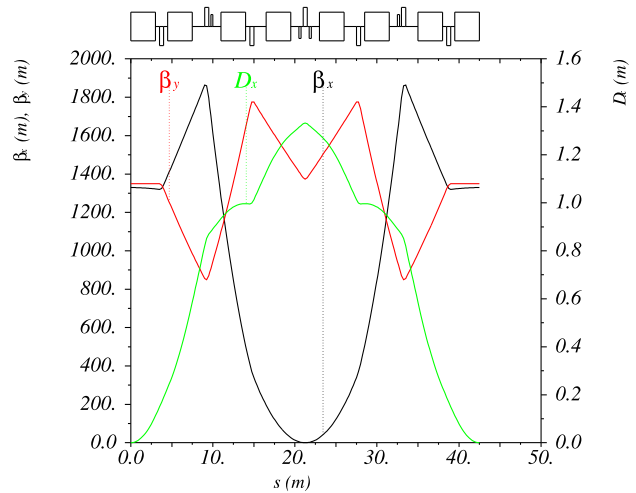}
\caption{CCB optics.} 
\label{fig:CCBoptics}
\end{figure} 

Since the chromaticity correction scheme is independent of particular BES and FFB implementations, 
we first focus on the CCB design. Figure~\ref{fig:CCBoptics} shows CCB optics developed 
for the MEIC ion collider ring following the concept illustrated in Fig.~\ref{fig:IRscheme2}. 
Note that, in comparison to the simplified schematic of Fig.~\ref{fig:IRscheme2}, a larger number 
of quadrupoles were used to better control the beta function behavior. The CCB is composed of 
eight symmetrically-arranged alternating bends with seven quadrupoles placed symmetrically between 
them. The sum of the absolute values of all dipole bending angles is 480~mrad, while, due to 
the alternation of their bending directions, the net CCB bending angle is 120~mrad.
The quadrupole strengths are adjusted to produce a total CCB transfer matrix that meets 
the symmetry requirements of Eq.~(\ref{eq:symm_cond}). 

\begin{figure}[t!]
\includegraphics[height=0.95\columnwidth,angle=-90]{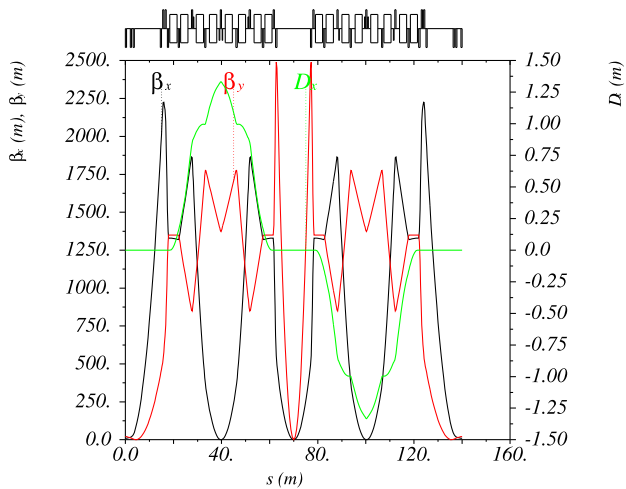}
\caption{Complete IR optics of the MEIC ion collider ring. The IP is located at the symmetry point 
in the middle.} 
\label{fig:completeIRoptics}
\end{figure} 

For the conceptual study presented in this paper, a simple quadrupole doublet is used 
for the FFB. As shown in Table~\ref{tab:MEICpar}, the design values of the horizontal and 
vertical $\beta$-functions at the IP $(\beta_{x,y}^*)$ are 10 and 2~cm, respectively. The 
distance from the IP to the front face of the nearest FFB quadrupole is assumed to be 7~m, 
which, in combinations with the $\beta_{x,y}^*$ values, determines the maximum values of the 
$\beta$-functions in the IR region. The FFB optics design also determines the Twiss function 
values at the entrance into the CCB. The FFB quadrupole strengths, lengths and spacing are 
adjusted so that the $\beta$-functions have roughly equal values at the entrance into the CCB 
and the $\alpha$-functions $(\alpha_{x,y} \equiv -(\partial \beta_{x,y}/\partial s)/2)$ 
are equal to zero in the middle of the CCB. A BES consisting of 7 quadrupoles then 
matches the CCB optics to the ring's regular lattice. This involves a substantial beam 
extension and therefore requires an appropriately large amount of longitudinal space. 
The IR is arranged symmetrically with respect to the IP. The complete IR optics of the MEIC 
ion collider ring is shown in Fig.~\ref{fig:completeIRoptics}. The detector solenoid and its coupling 
compensation elements are not included in this design. Note that the CCB bends on the opposite sides 
of the IP are reversed making the dispersion anti-symmetric with respect to the IP. This is also done 
for the purpose of electron and ion IR footprint matching.

The optics shown in Figs.~\ref{fig:CCBoptics} and \ref{fig:completeIRoptics} corresponds to the 
collider mode of operation. Since the injected beam's emittance is typically much greater than 
after acceleration and cooling, keeping the collider-mode optics at injection may result in an 
unacceptably large beam size. Therefore, to keep the magnet apertures reasonable, the maximum 
$\beta$-functions occurring in the IR section may need to be reduced at injection and during 
acceleration and then restored for the collider operation. The described IR design allows a 
straightforward implementation of such a ``$\beta$ squeeze''. Since entering and exiting the CCB 
the dispersion is suppressed and the $\alpha$-functions 
are almost zero, the CCB optics is compatible with any initial $\beta$-functions without any 
adjustment required as long as $\alpha_{x,y}$ have appropriate close-to-zero values. Thus, 
the size of the $\beta$-functions inside the IR can be controlled by adjusting the $\beta$-function 
values at the end of the BES. This requires only changing the optics of the BES with the rest of 
the ring optics intact.

\section{Momentum acceptance and dynamic aperture} 

\subsection{Chromaticity compensation and momentum acceptance} 
In accordance with our chromaticity compensation concept, two sextupole pairs are inserted 
in each CCB. The sextupoles in each pair are identical and are placed symmetrically with 
respect to the center of the CCB. The sextupoles are shown in Figs.~\ref{fig:CCBoptics} 
and \ref{fig:completeIRoptics} by the shorter bars. The sextupole positions are chosen 
at the points where the dispersion is near maximum and there is a large difference between 
the horizontal and vertical $\beta$-functions. The two parameters corresponding to the 
strengths of the sextupole pairs are adjusted to compensate the horizontal $\xi_x$ and 
vertical $\xi_y$ natural linear chromaticities.

\begin{figure}[t!]
\includegraphics[width=0.95\columnwidth]{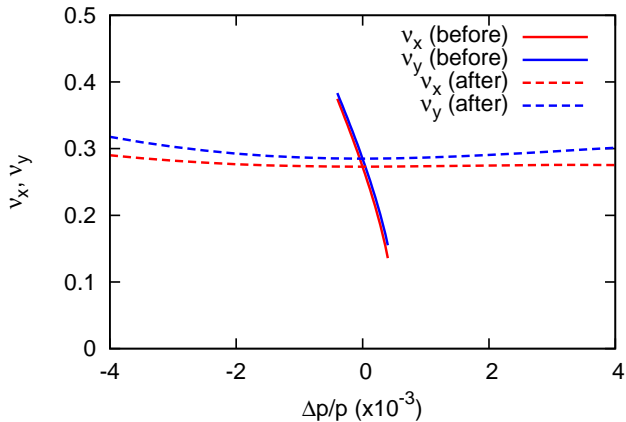}
\caption{Momentum dependence of the horizontal $\nu_x$ (red) and vertical $\nu_y$ 
(blue) fractional betatron tunes before (solid lines) and after 
(dashed lines) the linear chromaticity compensation.} 
\label{fig:chrom_dep_ccb}
\end{figure}

In this collider ring design~\cite{des_rep12}, the total values of 
$\xi_x$ and $\xi_y$ are -278 and -268, respectively. Contributions of the IRs to 
$\xi_x$ and $\xi_y$ are -254 (91.4\%) and 245 (91.3\%), respectively. Therefore, 
contribution of the remainder of the ring is almost negligible and can also be 
compensated locally by the CCBs. The two sextupole families are used to adjust the 
slopes of the chromatic horizontal and vertical betatron tune curves to zero. 
The chromatic tune dependence before and after the compensation is shown in 
Fig.~\ref{fig:chrom_dep_ccb}. The horizontal and vertical tune variations are 
less than 0.02 and 0.03, respectively, within $\Delta p/p$ of about $\pm 4 \times 10^{-3}$.

\begin{figure}[b!]
\includegraphics[height=0.95\columnwidth,angle=-90]{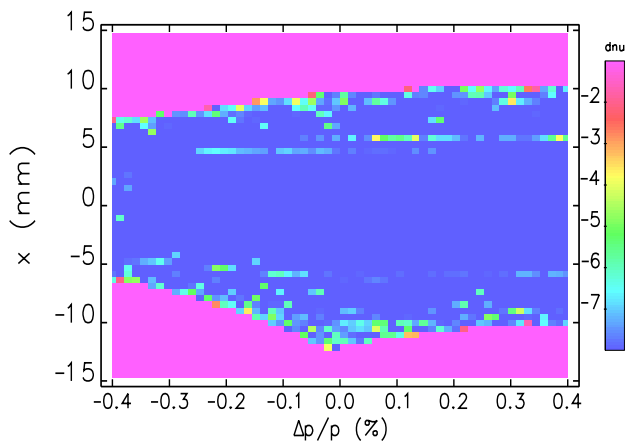}
\caption{Frequency map in the $(x-\Delta p/p)$ phase space. The color reflects the tune change in 
terms of the tune diffusion $d$ defined as $d \equiv \log(\Delta\nu_x^2+\Delta\nu_y^2)$.} 
\label{fig:FreqMap_ccb}
\end{figure} 

An effective method to evaluate the nonlinearity of a particle's dynamics is to obtain 
a frequency map~\cite{laskar95} from particle tracking. At the design 60 GeV/$c$ proton 
beam momentum of the MEIC ion collider ring, the maximum horizontal and vertical rms beam 
sizes are 3.2 and 1.6~mm, respectively, due to a relatively large detector space of $\pm 7$~m. 
This makes it more challenging to obtain a sufficiently large dynamic aperture in the horizontal 
dimension. Therefore, the frequency map is first computed in the $(x-\Delta p/p)$ phase space as 
shown in Fig.~\ref{fig:FreqMap_ccb}. The map is obtained by tracking particles for 2000 turns using 
a well-established accelerator simulation software {\it Elegant}~\cite{borland_elegant}. 
All ring components are modeled as canonical kick elements with exact Hamiltonians retaining 
all orders in momentum offset. The magnet fields are approximated as hard-edge and the lattice 
is assumed perfect, i.e. containing no alignment or field errors. The simulation does not 
take into account effects such as beam-beam, intra-beam scattering (IBS), etc. 
The particles are launched parallel to the beam axis at the entrance into one of the CCBs 
where the horizontal and vertical $\beta$-functions are about 1330 and 1350~m, respectively. 
The color in Fig.~\ref{fig:FreqMap_ccb} reflects the tune change in terms of the tune diffusion $d$ 
defined as $d \equiv \log(\Delta\nu_x^2+\Delta\nu_y^2)$ where $\Delta\nu_{x,y}$ is the tune 
change from the first to the second half of the simulation in the horizontal and vertical planes, 
respectively. The diffusion index $d$ is one of the best criteria~\cite{steier10} to determine 
the particle's long term stability and enables one to see the nonlinear behavior, i.e. the 
more negative the diffusion, the smaller the nonlinear effects are.

Two conclusions can be drawn immediately from the frequency map in Fig.~\ref{fig:FreqMap_ccb}:
\begin{enumerate}
\item 
the momentum acceptance can easily reach $\Delta p/p$ of $\pm 4 \times 10^{-3}$ with only 
the linear chromaticity compensation,
\item 
the uniform color distribution means that there are no strong resonant perturbations in 
the particle tracking.
\end{enumerate}

The simulation was terminated at $\Delta p/p = \pm 4 \times 10^{-3}$, which is about 
$\pm 14 \sigma_\delta$ (an rms momentum spread $\sigma_\delta$ of about $3 \times 10^{-4}$ is 
expected after cooling) and demonstrates a sufficient momentum acceptance. Thus, without 
any compensation of higher-order chromaticities, the symmetry concept immediately results in 
an excellent momentum acceptance. Figure~\ref{fig:FreqMap_ccb} also indicates that the horizontal 
dynamic aperture size is about $\pm 10$~mm, which is reasonable given the large compensated 
values of the natural chromaticities and the fact that there was no nonlinear optimization 
after the linear chromaticity compensation. However, due to the large beam extension required 
to achieve the ambitiously small $\beta^*$ values, clearly, further optimization of the 
dynamic aperture using multiple sextupole and octupole families is required. One approach 
to such optimization is presented in Section~\ref{sec:da_ccb}.

\begin{figure}[b!]
\includegraphics[height=0.95\columnwidth,angle=-90]{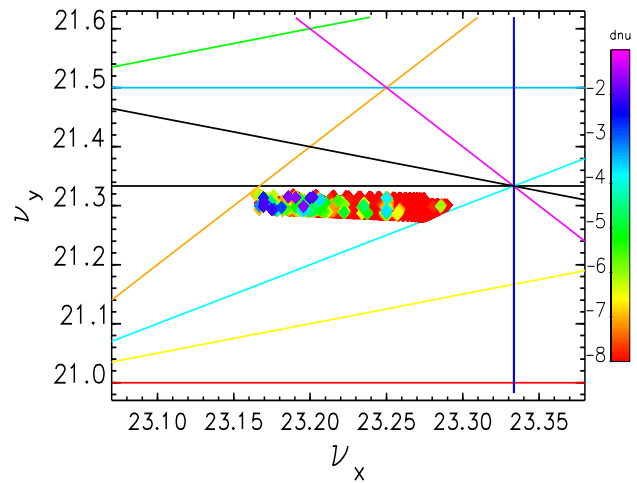}
\caption{Betatron tune footprint with the color indicating the tune diffusion $d$.} 
\label{fig:tunefp_ccb}
\end{figure} 

Figure~\ref{fig:tunefp_ccb} shows the tune footprint corresponding to the same tracking as 
the frequency map in Fig.~\ref{fig:FreqMap_ccb}. As with the frequency map, the color of the 
tune footprint reflects the tune change in terms of the diffusion index. The lines indicate 
betatron resonances up to the 3rd order; higher-order resonances are not shown in this plot. 
Since the tunes in Fig.~\ref{fig:tunefp_ccb} are calculated for particles with initial 
coordinates in the $(x-\Delta p/p)$ phase space, the vertical tune variation arises only 
from the chromatic tune dependence, which is around 0.03 within $\Delta p/p$ of 
$\pm 4 \times 10^{-3}$ as shown in Fig.~\ref{fig:chrom_dep_ccb}. The horizontal tune 
variation is about 0.14, which is significantly greater than the chromatic tune change of 0.02 
corresponding to $\Delta p/p$ of $\pm 4 \times 10^{-3}$ in Fig.~\ref{fig:chrom_dep_ccb}. 
This indicates that, after the linear chromaticity compensation, the horizontal betatron 
motion produces a substantial tune change, which can drive particles close to or into a resonance 
mode and cause particle loss. The tune change due to the particle's transverse motion is 
discussed in detail in Section~\ref{sec:da_ccb}. Overall, the tune footprint immediately 
reveals the distribution of the particle tunes, which allows one to understand the tune trend 
due to the particle motion and subsequently adjust the design tunes to avoid approaching or 
crossing strong resonances.

\begin{figure}[t!]
\includegraphics[height=0.95\columnwidth,angle=-90]{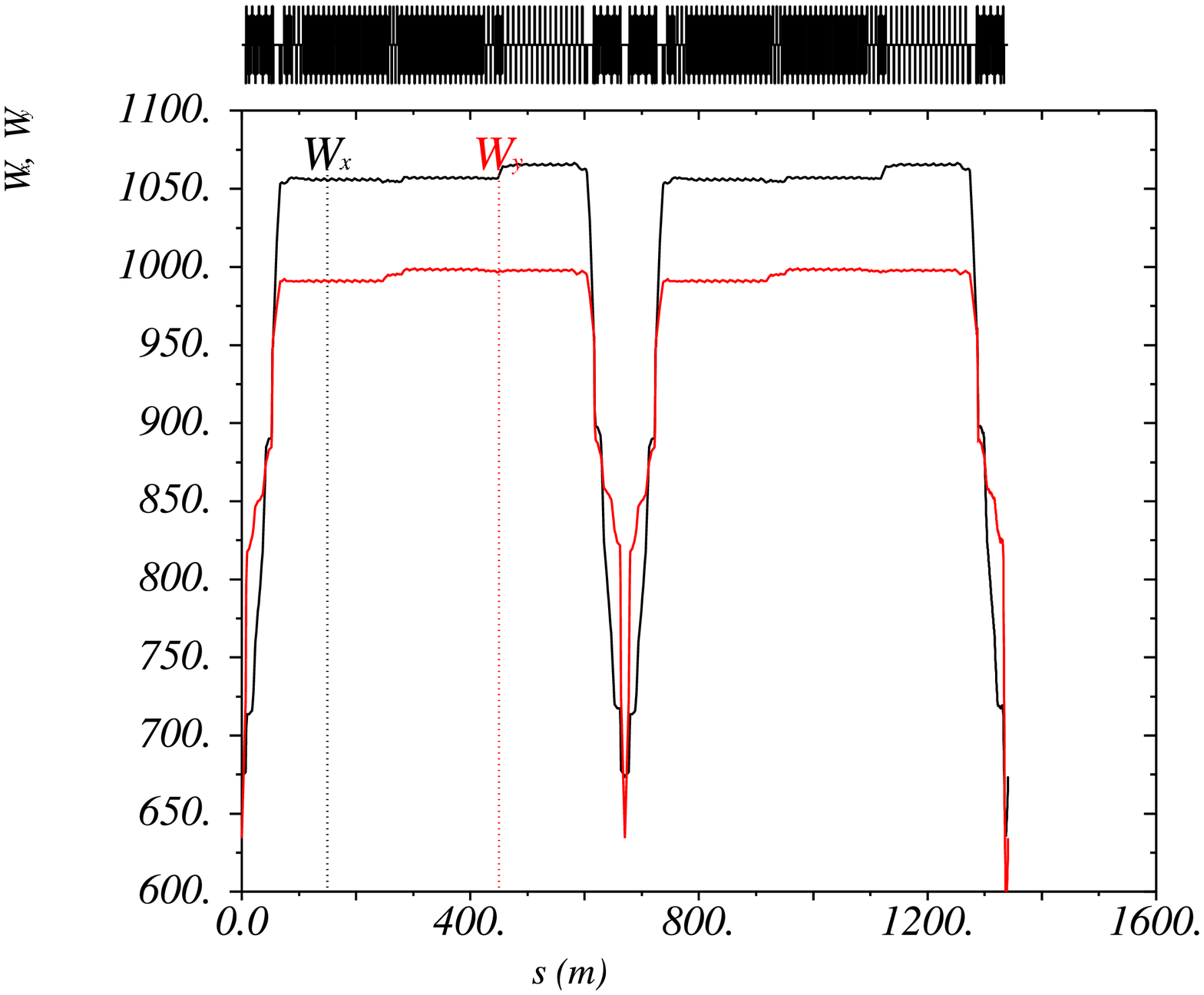}
\includegraphics[height=0.95\columnwidth,angle=-90]{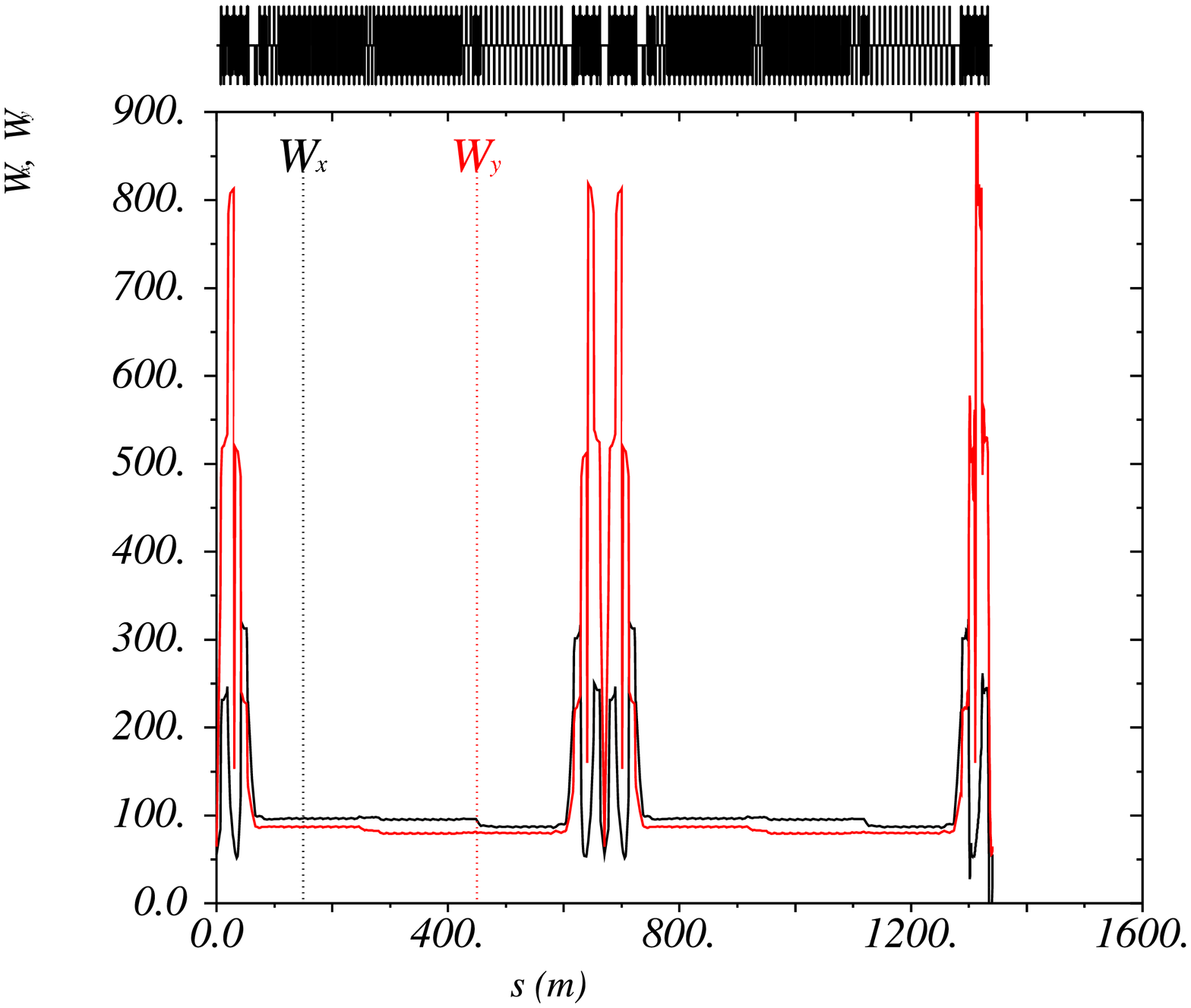}
\caption{Horizontal and vertical chromatic amplitude functions $W_{x,y}$ plotted over the whole 
ion collider ring length before (top) and after (bottom) the linear chromaticity compensation.} 
\label{fig:W_ccb}
\end{figure}

An important feature of our symmetry concept is compensation of the chromatic beam smear at the IP. 
It is attained together with the linear chromaticity correction when two CCBs are placed 
symmetrically around the IP. One way to demonstrate this is by using the chromatic amplitude 
functions $W_{x,y}$, which are directly related to the chromatic derivatives of the 
$\beta$-functions $\partial \beta_{x,y}/\partial \delta$~\cite{wfunc_madx}. 
Figure~\ref{fig:W_ccb} shows the $W_{x,y}$ functions plotted over the whole ion collider 
ring length before and after the linear chromaticity correction, respectively. The fact 
that $W_{x,y}$ are close to zero at the IP indicates that the chromatic beam smear is compensated.

\subsection{Comparison to the conventional distributed-sextupole scheme}
To demonstrate the advantages of the proposed symmetry concept, we compared its performance 
with that of a more conventional ``distributed-sextupole'' (DS) chromaticity compensation scheme 
based on introducing two sextupole families in the arc FODO cells. To make a fair comparison 
to the CCB scheme, the CCBs were taken out of the ring optics to eliminate their chromatic 
contribution, and the BESs were connected directly to the FFBs. Additional FODO cells and 
two more sextupole families were introduced in the regions between the arcs and BESs for 
higher-order chromaticity correction. The phase advances between the sextupoles and between 
the sextupoles and IP were adjusted to cancel the sextupole nonlinear kicks and to optimize 
the 2nd-order chromaticity compensation as discussed in~\cite{nosochkov92}. These modifications 
to the linear optics resulted, in particular, in higher betatron tunes but the initial 
horizontal $\xi_x$ and $\xi_y$ vertical natural chromaticities of -199 and -262, respectively, 
were still smaller than in the CCB scheme.

First, for the linear chromaticity compensation, two sextupole families were introduced in 
the arcs. Following the same guidelines as in the CCB case, the sextupoles were placed at 
locations with nearly maximum dispersion and a large difference between the horizontal and 
vertical $\beta$-functions. With 52 FODO cells in the two arcs of the ion collider ring, one 
might expect that such a globally distributed sextupole scheme can effectively reduce the 
sextupole strengths with a consequent suppression of the nonlinear effects introduced by 
the sextupole fields. However, the $\beta$-functions of a regular arc FODO cell are much 
smaller than those in the CCBs. Hence, even though the DS scheme has a total of 
104 sextupoles compared to only 16 sextupoles in the CCB scheme, the sextupole strengths in 
the DS scheme are still much stronger than those in the CCB scheme. 
Table~\ref{tab:sxt_par} summarizes the parameters of the two different linear chromaticity 
correction schemes including the values of the optical functions at the sextupole locations 
and the sextupole strengths.

\begin{table}[t]
\caption{Comparison of the optical functions at the sextupole locations and of the sextupole 
strengths $\partial^2 B_y/\partial x^2$ for 60~GeV/$c$ protons in cases of the DS and 
CCB-based linear chromaticity correction schemes.}
\begin{ruledtabular}
\begin{tabular}{lcccc}
Parameter & \multicolumn{2}{c}{DS scheme}  & \multicolumn{2}{c}{CCB scheme} \\ 
& \multicolumn{4}{c}{Sextupole family} \\ 
& 1 & 2 & 1 & 2 \\ \hline
\# of sextupoles & 52 & 52 & 8 & 8 \\
$\beta_x$ (m) & 13.8 & 6.0 & 1614 & 2.3 \\
$\beta_y$ (m) & 6.0 & 13.8 & 944 & 1396 \\
$D_x$ (m) & 1.19 & 0.8 & 0.9 & 1.3 \\
$\partial^2 B_y/\partial x^2$ (T/m$^2$) & 4000 & -7000 & 238 & -288 \\
\end{tabular}
\end{ruledtabular}
\label{tab:sxt_par}
\end{table}

\begin{figure}[b!]
\includegraphics[width=0.95\columnwidth]{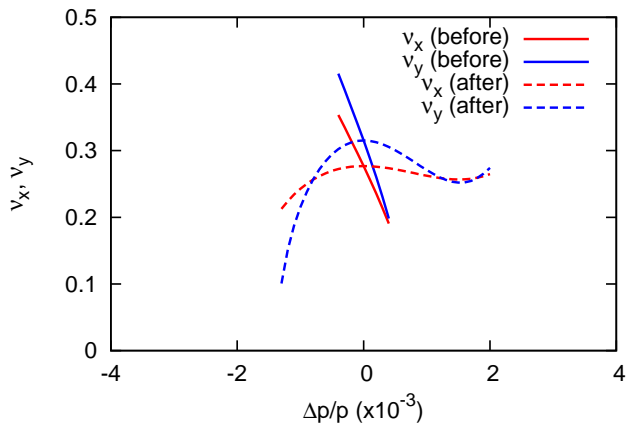}
\caption{Chromatic dependence of the fractional betatron tunes before and after 
the linear chromaticity compensation using the DS approach.} 
\label{fig:chrom_dep_ds_2sxt}
\end{figure} 

Figure~\ref{fig:chrom_dep_ds_2sxt} compares the chromatic dependence of the fractional betatron 
tunes before and after the linear chromaticity compensation using the two sextupole families in 
the arc FODO cells. The slopes of the horizontal $\nu_x$ and vertical $\nu_y$ chromatic betatron 
tune curves are zero at $\Delta p/p = 0$ after the correction, which reduces the tune variation 
and somewhat extends the momentum acceptance. However, the second derivatives of the corrected 
chromatic curves (the 2nd-order chromaticities) start dominating the tune spread and still limit 
the momentum acceptance. Therefore, we introduce two additional families to a total of 128 sextupoles 
close to the IR to compensate the 2nd-order chromaticities. For the optimal compensation, 
we set the phase advance between the sextupoles and IP following the approach described 
in~\cite{nosochkov92}. The strengths of the new sextupole families were comparable to those of 
the two original families in Table~\ref{tab:sxt_par}.

\begin{figure}[t!]
\includegraphics[width=0.95\columnwidth]{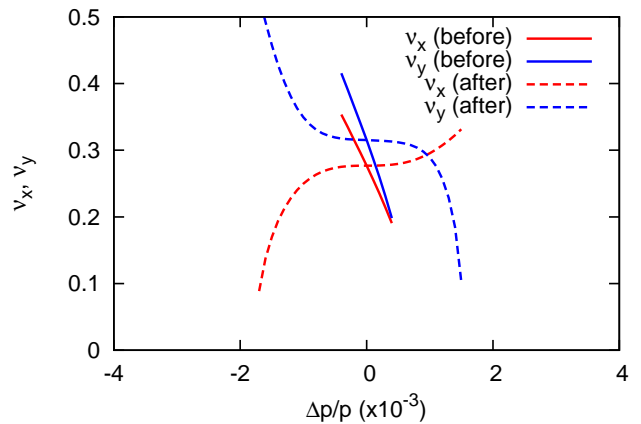}
\caption{Chromatic dependence of the fractional betatron tunes before and after the chromaticity 
compensation up to 2nd order.} 
\label{fig:chrom_dep_ds_4sxt}
\end{figure} 

\begin{figure}[b!]
\includegraphics[height=0.95\columnwidth,angle=-90]{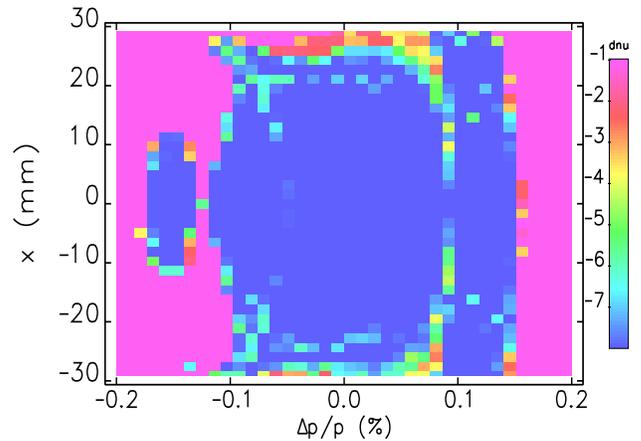}
\caption{Frequency map in the $(x-\Delta p/p)$ phase space for the 2nd-order DS scheme with 
the color indicating the tune diffusion $d$.} 
\label{fig:freqmap_ds}
\end{figure}

\begin{figure}[t!]
\includegraphics[height=0.95\columnwidth,angle=-90]{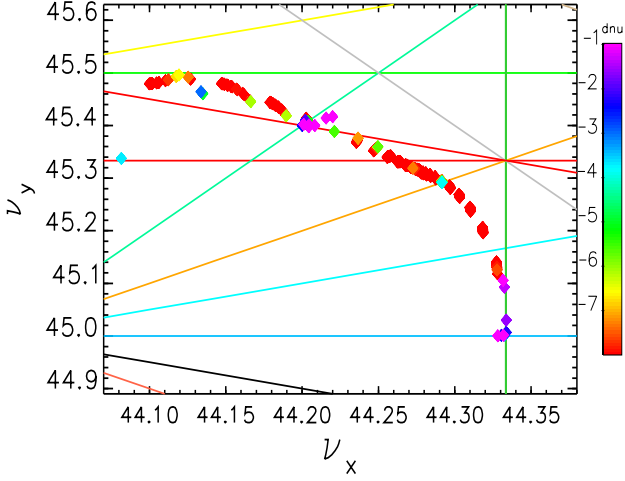}
\caption{Tune footprint for the 2nd-order DS scheme with the color indicating the tune diffusion $d$.} 
\label{fig:tunefp_ds}
\end{figure}

Figure~\ref{fig:chrom_dep_ds_4sxt} shows the chromatic tune dependence after combining the linear 
and 2nd-order chromaticity compensations. Figure~\ref{fig:chrom_dep_ds_4sxt} indicates that the 
momentum acceptance is extended further in comparison to Fig.~\ref{fig:chrom_dep_ds_2sxt}. 
However, the strong sextupole fields give rise to even higher-order chromaticities still leading 
to an impractically large tune spread. The frequency map and tune footprint for this 2nd-order DS 
chromaticity compensation scheme are also calculated using particle tracking and plotted in 
Figs.~\ref{fig:freqmap_ds} and \ref{fig:tunefp_ds}, respectively. Because of such a large chromatic 
tune spread, the particles experience numerous resonances, which limit the momentum acceptance.

\begin{figure}[b!]
\includegraphics[height=0.95\columnwidth,angle=-90]{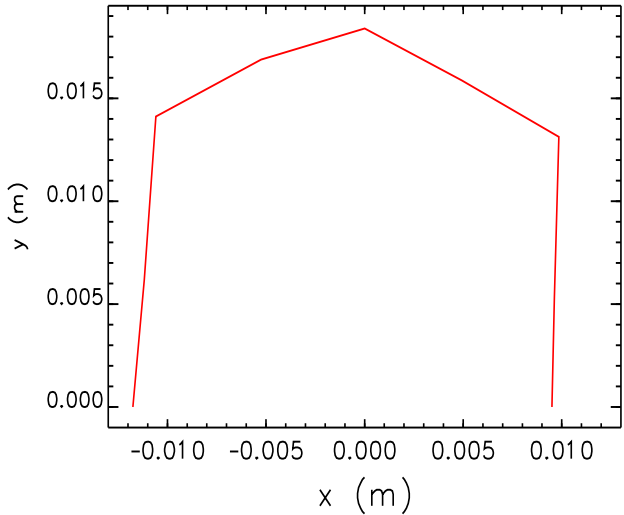}
\caption{Dynamic aperture in the $(y-x)$ space at the entrance into one of the CCBs 
for the on-momentum particle.} 
\label{fig:da_nooct}
\end{figure} 

\subsection{Dynamic aperture with the CCB scheme}
\label{sec:da_ccb}
A ring's property that is equally important to its momentum acceptance is its dynamic aperture (DA), 
i.e. a region in the transverse plane of stable particle motion. Therefore, we study the 
ion collider ring's DA for the case of the CCB-based chromaticity compensation scheme. The DA 
is explored by tracking particles for 1000~turns while progressively increasing their initial 
horizontal $(x)$ and vertical $(y)$ amplitudes until the boundary between survival and loss is found. 
Figure~\ref{fig:da_nooct} shows the on-momentum particle's DA at the entrance into one of the CCBs.  
The DA reaches $\sim 10$~mm horizontally and $\sim 18$~mm vertically, which is quite reasonable 
given the large natural chromaticities and the fact that there was no nonlinear optimization 
after the linear chromaticity correction. However, due to the large beam extension required to 
achieve the ambitiously small $\beta_{x,y}^*$ values at the MEIC (see Table~\ref{tab:MEICpar}), 
these horizontal and vertical DA sizes correspond to only $\sim 4\sigma_x$ and 
$\sim 15\sigma_y$, respectively. Thus, optimization of the DA using multiple sextupole and octupole 
families is required.

\begin{figure}[b!]
\includegraphics[height=0.95\columnwidth,angle=-90]{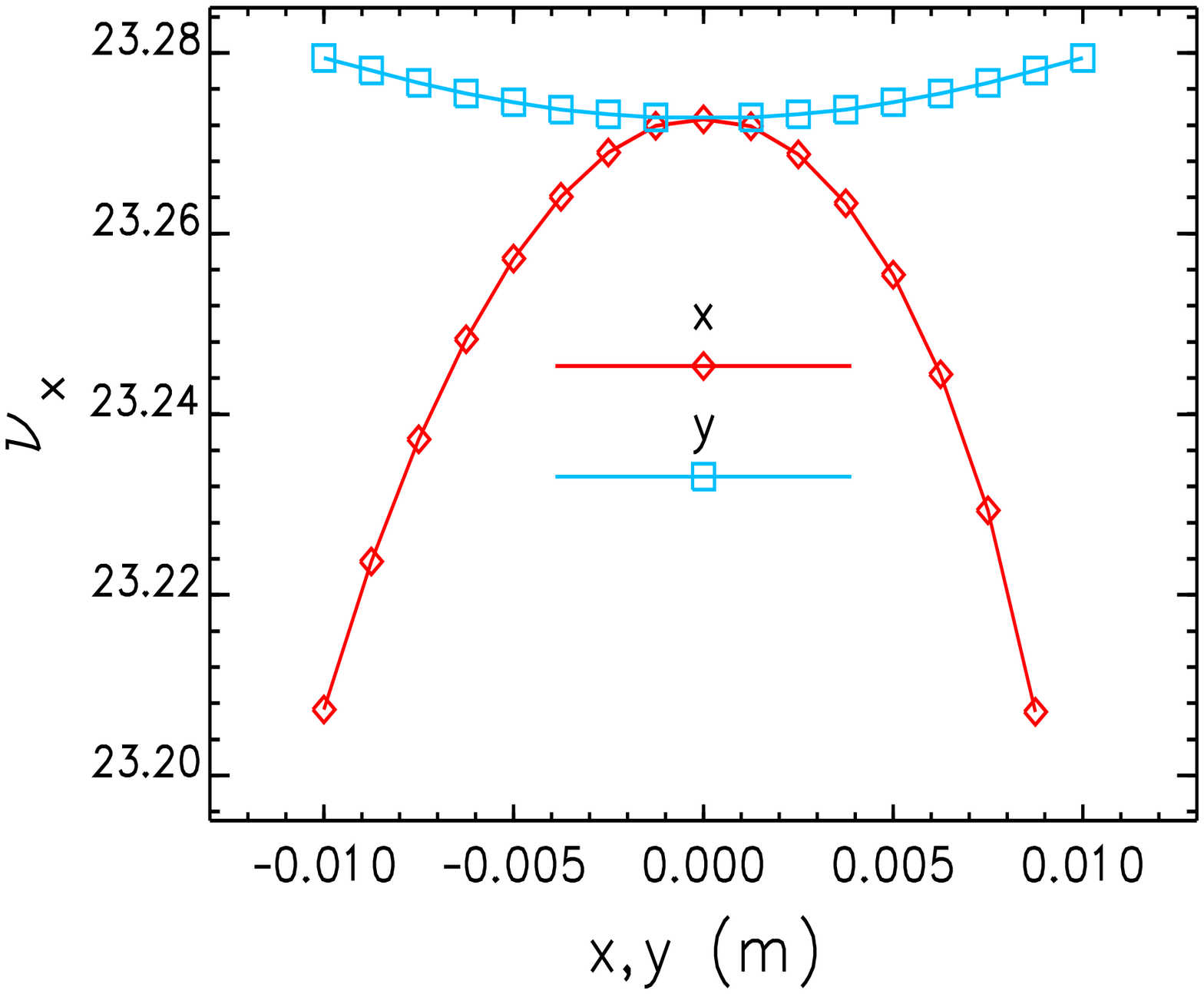}
\includegraphics[height=0.95\columnwidth,angle=-90]{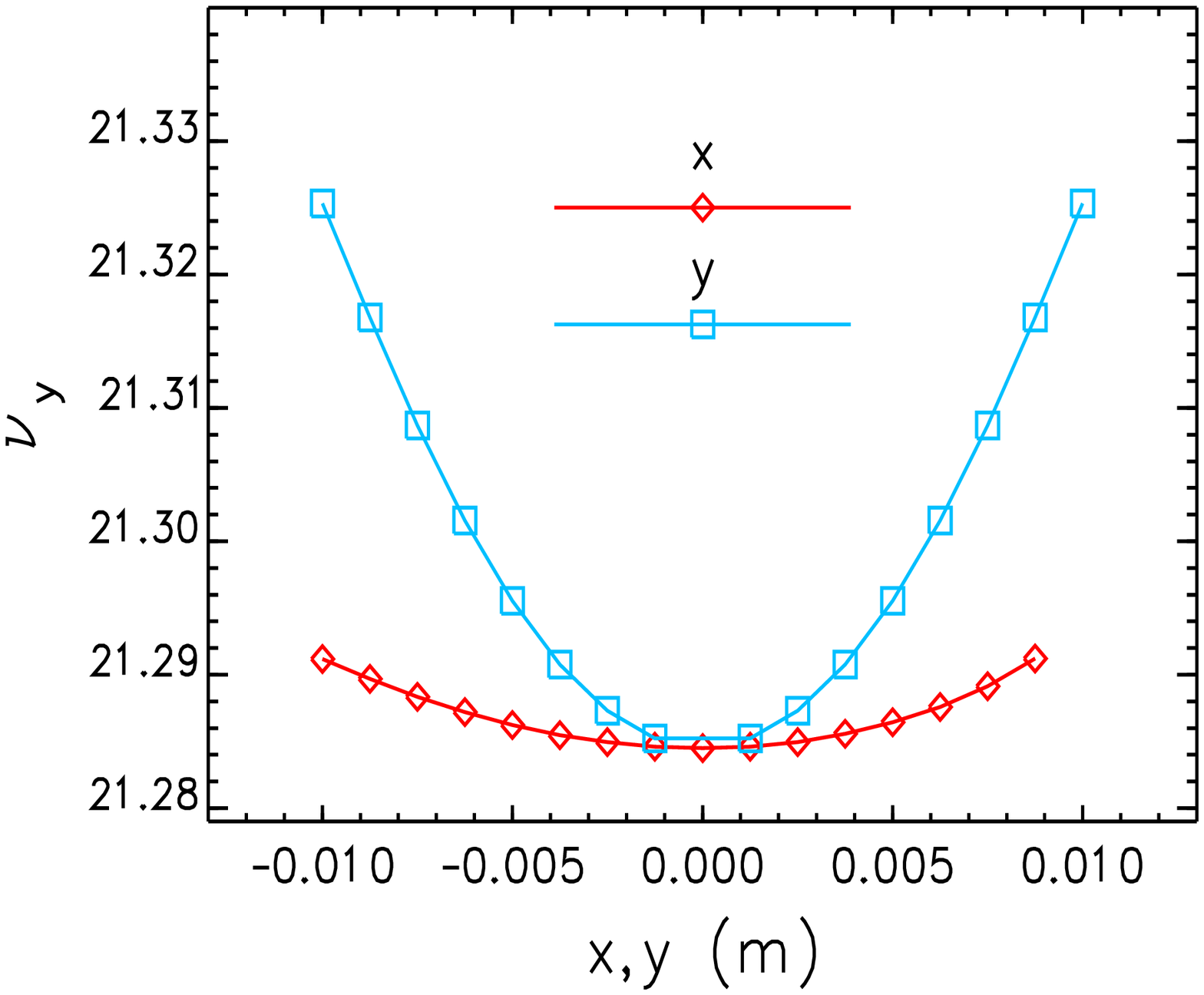}
\caption{Dependence of the horizontal $\nu_x$ (top) and vertical $\nu_y$ (bottom) betatron tunes 
on the initial transverse amplitudes $x$ and $y$.} 
\label{fig:amptunedep_nooct}
\end{figure}

For the on-momentum particle, there is clearly no tune variation due to the chromatic tune 
dependence. The tune change comes from the nonlinearity of the transverse betatron motion caused 
by the nonlinear fields, in particular, of the sextupoles. Thus, we explore the dependence of the 
betatron tunes on the transverse motion. For a particle with initial transverse amplitudes $x$ and 
$y$, its horizontal $\nu_x$ and vertical $\nu_y$ tunes are calculated by tracking it for a number 
of turns. The obtained tune dependence is shown in Fig.~\ref{fig:amptunedep_nooct}: $\nu_x$ is 
plotted as a function of $x$ and $y$ at the top while $\nu_y$ is plotted as a function of $x$ 
and $y$ at the bottom.

The simulations show that, after the sextupole compensation, there is a large amplitude-dependent 
tune shift. This is due to the fact the nonlinearity of the sextupole fields causes stronger 
focusing of the particles with larger transverse amplitudes. As seen from Fig.~\ref{fig:amptunedep_nooct}, 
$\nu_x$ and $\nu_y$ change from the design values by 0.08 (the red line in the top plot) and 0.04 
(the blue line in the bottom plot), respectively, within $x$ and $y$ of about $\pm 10$~mm. 
Such a large tune deviation can easily drive particles to approach or cross betatron resonances, 
which inevitably causes their loss. The tune shift with amplitude limits the ring's DA, especially 
in the horizontal direction due to the large horizontal beam extension in the ion collider ring 
(the maximum horizontal rms beam size $\sigma_x$ along the ring is 3.2~mm). 
Figure~\ref{fig:amptunedep_nooct} also shows that there is very little coupling between 
the two transverse dimensions: $\nu_x$ is almost independent of the vertical motion (the blue line 
in the top plot of Fig.~\ref{fig:amptunedep_nooct}) while $\nu_y$ is unaffected by the horizontal 
motion (the red line in the bottom plot of Fig.~\ref{fig:amptunedep_nooct}).

\begin{figure}[t!]
\includegraphics[height=0.95\columnwidth,angle=-90]{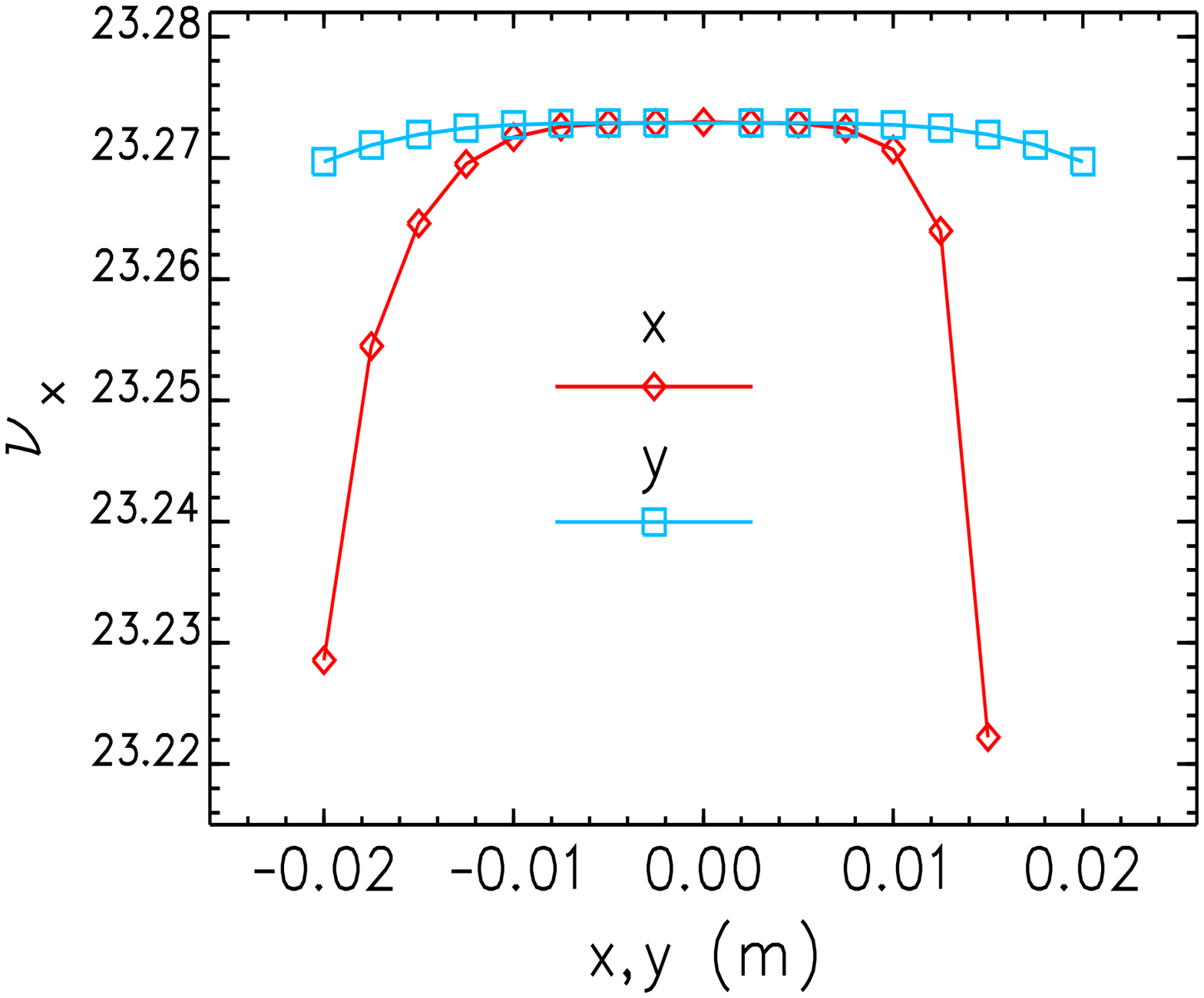}
\includegraphics[height=0.95\columnwidth,angle=-90]{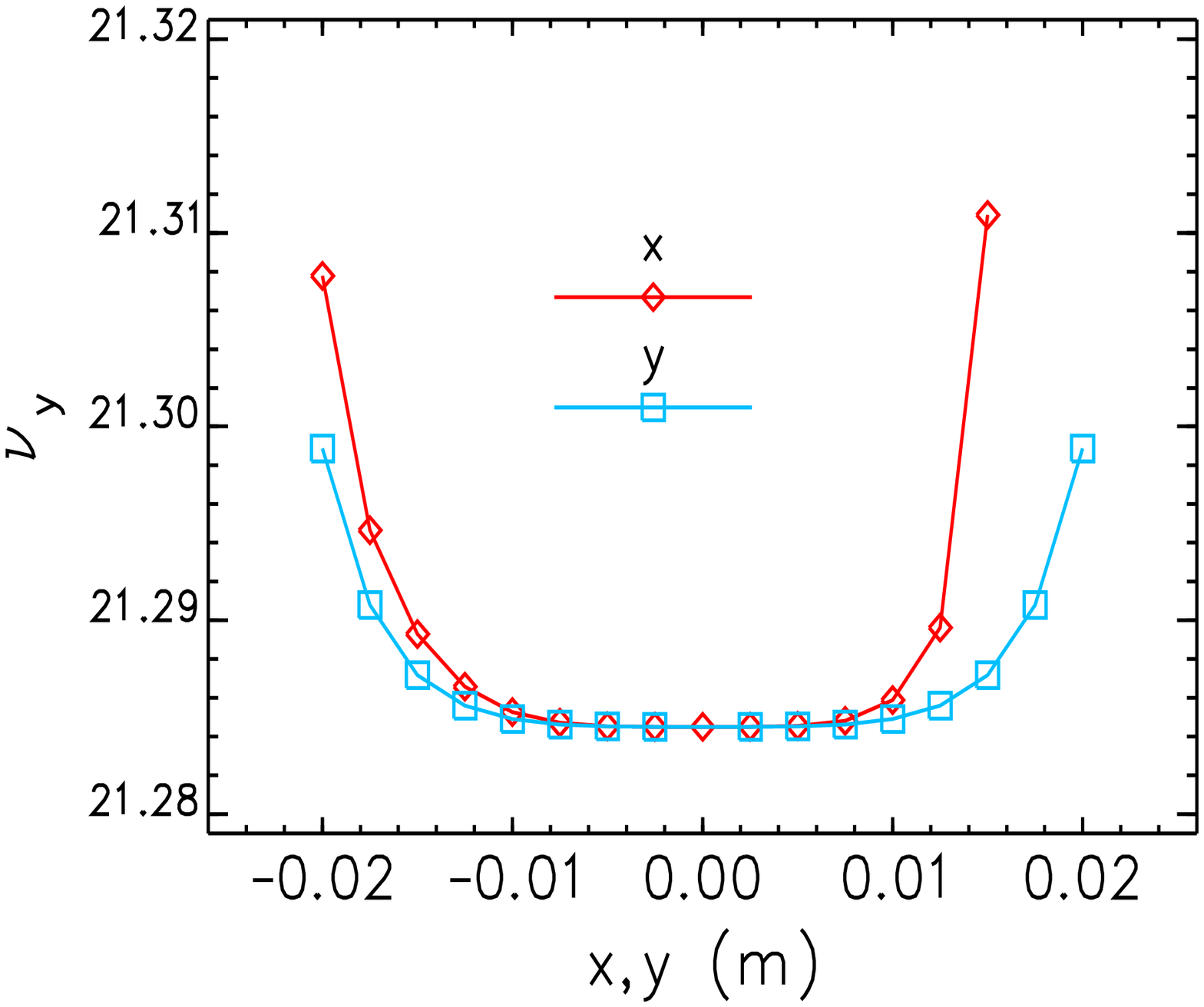}
\caption{$\nu_x$ (top) and $\nu_y$ (bottom) as functions of $x$ and $y$ after compensation of the 
1st-order amplitude dependent tune shifts.} 
\label{fig:amptunedep_withoct}
\end{figure}

\begin{figure}[t!]
\includegraphics[height=\columnwidth,angle=-90]{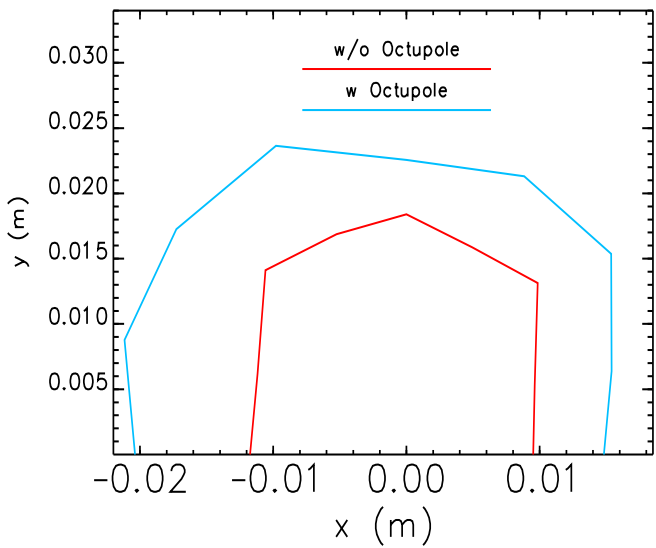}
\caption{Dynamic aperture in the $(y-x)$ space at the entrance into one of the CCBs 
for the on-momentum particle without (red) and with (blue) the octupole compensation of the 
1st-order amplitude-dependent tune shift.} 
\label{fig:da_withoct}
\end{figure}

A straightforward mechanism to compensate the amplitude-dependent tune shift caused by the 
sextupoles is to introduce octupoles. Depending on the octupole locations, both the amplitude 
dependent tune shift and higher-order chromaticities can be controlled by adjusting the octupole 
strengths. Since the momentum acceptance is large enough after the linear chromaticity correction 
with the CCBs, we focus on compensation of the amplitude-dependent tune shift. Therefore, octupole 
families should be placed in dispersion-free regions to leave the chromatic correction unaffected. 
Also, to reduce the required octupole strengths, they should be placed at large $\beta$-function 
points. Following these considerations, three families of octupoles are introduced in the BES and 
FFB regions. It was the most convenient for us to find a solution for the octupole strengths that 
provide compensation of the 1st-order amplitude dependent tune shifts, using the arbitrary-order 
expansion tools of COSY Infinity~\cite{berz_cosy}. After this compensation, the horizontal and 
vertical tune variations are again obtained as functions of the initial transverse amplitudes $x$ 
and $y$ from particle tracking and are plotted in Fig.~\ref{fig:amptunedep_withoct}.
Compared to the tune changes of 0.08 and 0.04 within the transverse amplitude of about 
$\pm 10$~mm in Fig.~\ref{fig:amptunedep_nooct}, the horizontal and vertical tune changes are now 
0.04 and 0.03, respectively, within an amplitude range of over $\pm 15$~mm. Consequently, the DA 
is increased as shown by the blue line in Fig.~\ref{fig:da_withoct} in comparison to the red line 
without the octupole optimization.

After the octupole compensation, the limitation on the DA comes from the effect of 
the 2nd-order amplitude-dependent tune shift, which starts dominating the tune behavior 
in Fig.~\ref{fig:amptunedep_withoct}. This provides a guideline that the DA can be further 
enhanced by considering and optimizing the 1st-, 2nd-, and even higher-order amplitude-dependent 
effects. One possibility for future studies is to use a genetic algorithm to optimize the DA 
directly by simultaneously adjusting multiple sextupole and octupole families~\cite{borland10}. 
Note that magnet errors are not included in the above simulations. Addressing the question 
of error tolerances is the subject of our future studies.

\section{Conclusions}
We proposed a new symmetry-based interaction region concept for a high-luminosity collider 
ring design. Such a symmetric design allows simultaneous compensation of (a) the betatron tune 
spreads due to the 1st-order chromaticities, (b) chromatic beam smear at the IP, i.e. chromatic 
spread of focal point, and (c) other 2nd-order aberrations to the particle trajectory at the IP 
and on average over one turn, such as the 2nd-order dispersion effect and sextupole-generated 
beam smear due to the beam size. According to this concept, two specially-designed symmetric 
Chromaticity Compensation Blocks (CCBs) are placed symmetrically around an IP. Each CCB induces 
an angle spread in the passing beam such that it cancels the chromatic kick of the final focusing 
quadrupoles. We developed an analytic description of this approach and explicitly formulated 
2nd-order aberration compensation conditions at the IP. We verified our concept by developing a 
specific interaction region design and applying our chromaticity compensation scheme to 
a challenging case of a high-chromaticity small-$\beta^*$ MEIC ion collider ring~\cite{des_rep12}. 
We then compared performance of our symmetric CCB approach with that of the conventional 
distributed-sextupole chromaticity compensation scheme. The CCB approach resulted in a much greater 
momentum acceptance with a smaller number of weaker sextupoles. The dynamic aperture after the 
chromaticity correction is reasonably large, especially given the large natural chromaticities. 
We identified the factors limiting the dynamic aperture, namely, the 1st- and higher-order 
amplitude-dependent tune shifts, and demonstrated an approach to dynamic aperture optimization 
using octupole families. Further optimization may be needed depending on the specific requirements. 
Evaluation of the error impact will be addressed in the future studies.

\section{Acknowledgments}
This work was supported in part by the U.S. Department of Energy Small Business Technology 
Transfer Grant DE-SC0006272. This manuscript has been authored by Jefferson Science Associates, 
LLC under Contract No. DE-AC05-06OR23177 with the U.S. Department of Energy. The United States 
Government retains and the publisher, by accepting the article for publication, acknowledges 
that the United States Government retains a non-exclusive, paid-up, irrevocable, world-wide 
license to publish or reproduce the published form of this manuscript, or allow others to do so, 
for United States Government purposes. We are grateful to C.M.~Ankenbrandt, S.A.~Bogacz, 
A.M.~Hutton, G.A.~Krafft, F.C.~Pilat, and T.J.~Satogata for their help and advice.

\end{document}